\documentclass{article}
\usepackage{arabtex}
\usepackage{utf8}
\setcode{utf8}
\usepackage{arxiv}
\usepackage[utf8]{inputenc} % allow utf-8 input
\usepackage[T1]{fontenc}    % use 8-bit T1 fonts
\usepackage{hyperref}       % hyperlinks
\usepackage{url}            % simple URL typesetting
\usepackage{booktabs}       % professional-quality tables
\usepackage{amsfonts}       % blackboard math symbols
\usepackage{nicefrac}       % compact symbols for 1/2, etc.
\usepackage{microtype}      % microtypography
\usepackage{lipsum}		% Can be removed after putting your text content
\usepackage{graphicx}
\usepackage{natbib}
\usepackage{doi}

\title{T-Curator: a trust based curation tool for LOD logs}

\author{ \href{https://orcid.org/0000-0002-3794-844X}{\includegraphics[scale=0.06]{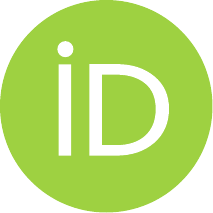}\hspace{1mm}Dihia LANASRI}\\
	ESI\\
	Algiers, Algeria\\	
	\texttt{ad\_lanasri@esi.dz} \\
}

\renewcommand{\shorttitle}{T-Curator for LOD Logs}

\hypersetup{
pdftitle={T-Curator: a trust based curation tool for LOD logs},
pdfsubject={NLP, ANALYTICS},
pdfauthor={D.LANASRI},
pdfkeywords={Linked Open Data , Query-logs , Data Curation , Tool},
}

\begin{document}
\maketitle

\begin{abstract}
Nowadays, companies are racing towards Linked Open Data (LOD) to improve their added value, but they are ignoring their SPARQL query logs. If well curated, these logs can present an asset for decision makers.
A naive and straightforward use of these logs is too risky because their provenance and quality are highly questionable. Users of these logs in a trusted way have to be assisted by providing them with in-depth knowledge of the whole LOD environment and tools to curate these logs. 
In this paper, we propose an interactive and intuitive trust based tool that can be used to curate these LOD logs before exploiting them. This tool is proposed to support our approach proposed in our previous work \cite{lanasri2020trust}.
\end{abstract}

% keywords can be removed

\section{Introduction}
Nowadays, an explosive fast growth of data is noticed. Linked open data are raised rapidly since their first publication; they started in May 2007 with only 12 interlinked datasets to exceed 1314 datasets in September 2023\footnote{\url{https://lod-cloud.net/}}. In a short period, LOD have become key sources for different communities, companies, organizations and governments to create and share knowledge and to increase their added value. Many LOD in different domains are published on the web in RDF format \cite{hogan2020knowledge}, like DBpedia \cite{lehmann2015dbpedia}, Nell \cite{mitchell2018never}, Yago \cite{hoffart2011yago2}, etc. DBpedia is one of the most famous open KG, composed of around 2 billion of RDF triples, with 27.2 million data links into external RDF datasets and it describes more than 400 million "facts"\footnote{\url{https://downloads.dbpedia.org/wiki-archive/data-set-39.html}}. 

This explosive growth, richness, variety and openness of LOD lead to growing interest to these sources. They represent a real important asset for individuals and companies. They are exploited in different manners \cite{lissandrini2020knowledge} Summarization, profiling, Exploratory search, Exploratory analysis \cite{berkani2020contribution,nebot2012building}, exploratory OLAP \cite{ravat2016enabling,colazzo2014rdf} and MD vocabularies like QB4OLAP \cite{etcheverry2012qb4olap}).

The wide exploitation of LOD datasets generates a large amount of SPARQL query logs. This second component of LOD environment, when open, represents a valuable and wealthy source of data. These logs are collected directly from SPARQL endpoints and published by some initiatives like LSQ\footnote{\url{https://aksw:github:io/LSQ/}} \cite{saleem2015lsq} and USEWOD\footnote{\url{http://usewod.org/}}. Or, they are provided by Question Answering (QA) systems \cite{rajpurkar2018know,diefenbach2018core} which generate in back-end SPARQL queries. Some users, who are not familiar with SPARQL syntax, can use different QA systems\footnote{\url{http://qa3.link/}} over LOD datasets to formulate their queries in natural Language then the system translates them into SPARQL queries. 

LOD logs have been used for different purposes such as statistical analysis \cite{bonifati2020analytical}, source selection \cite{tian2011enhancing} and multidimensional exploration \cite{khouri2019loglinc}. However, these logs present serious trust issues making their direct use, without any curation, risky; because they are usually published by unknown and less credible users with many intentions, who may provide inaccurate queries. Moreover, their expertise level may affect strongly their quality. 

Trust is a very complex concept linked to other crucial concepts such as Risk \cite{amaral2019towards}, Quality \cite{ceolin2015linking}, Provenance \cite{suriarachchi2016crossing} and Value \cite{sales2018common}. It is defined as \textit{“the subjective probability with which an agent expects that another agent or group of agents will perform a particular action on which its welfare depends”} \cite{gambetta2000can}. This definition of trust is the widely accepted one in the literature \cite{amaral2019towards}.

To achieving the curation of data by filtering, transforming and cleaning undesired data, Extract-Transform-Load (ETL) or data curation solutions are used. They play a key role in data preparation by leveraging traditional ETL operators and considering others related to trust. Data curation is a time and resource consuming task \cite{rezig2019data}, it may be manual, automatic or crowdsourced \cite{chen2020building}.

Management of trust in LOD logs is required, to achieve this goal, we define in this paper, which extend our previous paper \cite{lanasri2020trust}, a trust based curation tool for LOD logs. Our curation tool based on two steps: i) A log profiling of SPARQL queries to analyze their structure, quality and provenance, ii) the definition of ETL operators adapted to trust context, when orchestrated, they form an ETL-pipeline. 
This tool was developed for data analysts and data scientists to decide how to curate their logs.

This paper is organized as follows: Section 2 describes our related work. Section 3 details our developed tool. Section 4 presents a demonstration of our tool. Section 5 concludes our paper.

\section{Related Work}
Our work is related to Trust, Data curation and LOD logs tools. In what follows we will review the main tools used for these purposes.

\subsection{Trust Tools}: Trust was studied for LOD datasets \cite{hartig2009querying} an extended RDF (tRDF) to represent trust value on RDF triples and a tSPARQL query language are proposed\footnote{http://trdf.sourceforge.net/tSPARQL.shtml}. A POC also is proposed \cite{mazzieri2004fuzzy} to represent trust with Fuzzy RDF which is a syntactic and semantic extension of RDF. TriQL.P is a general purpose RDF browser that supports users in exploring RDF datasets, information can be filtered using a wide range of trust policies \cite{bizer2005triql}. The metaK\footnote{http://isweb.uni-koblenz.de/Research/MetaKnowledge} prototype is used to represent trust in SW with Named Graph \cite{dividino2009querying}.

\subsection{Data Curation Tools}: Many curation tools were proposed to clean data.  ETL tools like SETL \cite{deb2020setlbi} used to treat RDF data for OLAP analysis. Some commercial tools like Microsoft SSIS\footnote{https://docs.microsoft.com/en-us/sql/integration-services/sql-server-integration-services?view=sql-server-ver15} and Talend\footnote{https://www.talend.com/fr/} are used for data transformation. To insure quality of data in data lakes some services are developed citing CoreKG \cite{beheshti2018corekg} presenting a service for data curation for linking, enriching and annotating data.  KAYAK \cite{maccioni2018kayak} supporting data scientists in the definition, execution and optimization of data preparation pipelines in a data lake. In social data, Datasynapse is also proposed to curate these data\footnote{https://github.com/unsw-cse-soc/datasynapse} \cite{beheshti2019datasynapse}].

\subsection{LOD logs tools}: Some studies proposed a statistical analysis of the content of LOD query-logs. They have provided tools supporting such analysis like DARQL \cite{bonifati2018darql} mainly used to discover the inherent characteristics of the SPARQL queries via GUIs, and SEMLEX \cite{mazumdar2011semlex} that proposes a semantic analysis of the contents of query-logs. Finally a multidimensional patterns exploration tool is proposed in \cite{lanasri2019crumbs4cube} to detect the different MD patterns (Dimensions, facts, measures, etc.) from LOD Logs.

As detailed above, there is no tool proposed to curate and treat LOD logs and taking in consideration the dimension of trust. This motivates our proposal.

\section{T-Curator Tool Architecture Overview}
LOD logs are rich sources to be considered for decision making. However, using them in their raw format is risky since they are generated by unknown users with various profiles. We relate risks of LOD logs to Provenance (who generates queries?) and Quality (how is the state of queries?).
\\
In order to use safely these queries, we proposed in our last work \cite{lanasri2020trust} a number of trust based operators for our curation approach. Some are inspired from classical ETL, others are specific to trust in LOD logs. These operators are orchestrated by data analyst to construct her Trust based curation pipeline serving to curate LOD logs and keeping more trusted queries.
\\
The proposed operators are organized into three categories: extract operators, transform operators and load operators. Each operator has a common signature: $operator_name(input_Queries): (TrustQ; UTrustQ; RateOfTrust)$ 
For example \\$Deduplicator(Queries) : (TrustQ; UTrustQ; RateOfTrust)$.
An LOD query log is given as a set of $QL$ files, each $QL$ is a set of SPARQL queries $Q$. SPARQL queries are defined for matching a defined subgraph of triples < S P O > in the queried RDF graph. For instance, the following example illustrates a SPARQL query of scholarly data LOD log used in our demonstration:

\scriptsize${SELECT DISTINCT ?pred ?author_url ?author_name}$\\
\scriptsize${WHERE {<uri/conquer-query> bibo:authorList ?authorList.}}$\\
\scriptsize${?authorList ?pred ?author_url .}$\\
\scriptsize${?author_url foaf:name ?author_name}$

To evaluate the effectiveness of our curation approach, we calculate two metrics after each T-Curator operation:

i) $RateOfTrust=\frac{QL - ||TrustQ||}{QL}$
Where $QL$ is number of input queries, And ||TrustQ|| number of trusted queries. 
ii) Number of: Untrusted queries and Trusted queries.

The proposed tool is used to curate LOD query logs and keep only the trusted queries by orchestrating the trust operators to form a suitable curation pipeline. Our system adopts three tiers architecture basing on the MVC frameworks. This solution is developed using Java \& Scala for parallel programming. Jena API (ARQ and Core libraries) is also used to deal with SPARQL queries. As illustrated in figure \ref{architecture}, the layers are given as follows:

\subsection{Presentation layer} 
It represents the GUI provided to the data analyst in order to build her trust-aware curation pipeline. This interface allows her to decide how to cook her logs by selecting and combining the operators proposed in zone (a, Fig.\ref{tool}). Once the needed operators are selected, the curation pipeline is graphically generated in zone (b, Fig.\ref{tool}), the operators are automatically reorganized in a logical manner according to a certain order. The data analyst does not have hand on the order of operators in the pipeline. When it is running, the results of curation are given in the zone (c, Fig.\ref{tool}) where some details are given: sample of treated queries, number of trusted and untrusted queries and the rate of trust after each operation. This layer is developed using JavaFX\footnote{\url{https://openjfx.io/}}  with Scene builder\footnote{\url{https://gluonhq.com/products/scene-builder/}}. The importance of this tool is its customization; the data analyst can at each step decide which type of query to keep. In Business/academic operator, she can decide to keep just business queries which can be considered trustier.

\subsection{Business Layer}: 
It is the core of our tool. It contains the necessary operations and methods used to curate LOD logs. As illustrated in Fig.\ref{architecture}, five modules compose this layer:
\begin{enumerate} 
\itemsep=0pt
\item Extraction module: containing extract and format converter operators. They allow extracting select or construct SPARQL queries from log files with their meta-data like: IP address, execution dateTime and response code. Then parsing them using UTF8 decoder.
\item Transformation module: containing transformation operators which are grouped into three groups: 

•	Single queries: queries are treated one by one.  Business/Academic query extractor helps, using WHOSIP, to select business queries generated by professionals or academic ones generated from academic institutions.  Vulnerable query eliminator is used to delete all vulnerable queries that are generated by IPs appearing in a database of blacklisted IPs.  Complexity filter is important to detect shapes and depths of queries using [9] solution, complex queries indicate generally an expert profile behind which allows filtering queries.  Syntactic \& Semantic correctors allow correcting wrong queries basing on a REGEX and the algorithm proposed by [3]. Analytic/standard query selector allows selecting standard queries or analytic ones containing aggregate functions that reflect an analysis aim.

•	Interacted queries: the interactions between queries should be considered to understand their behavior. Robot query cleaner is used to discard all bot queries not generated by humans. Basing on the results of complexity analysis and the behavior of user to enhance her query, Expertise filter identifies expert from beginner or intermediate profiles. Deduplicator is used to keep unique queries and discard duplications while Topic clustering and Schema ranking are used to detect the topic of a given query basing on a created reference base and then deduplicate queries basing on the similarity of their triples <S P O>.

•	Interacted logs: the interaction between many LOD logs is considered by detecting semantic similarity between queries of different logs then regrouping them via Logs join operator. As each log is associated to a given source, Logs enrichment is used to detect the queries executed against this data source in other logs.

\item Trust annotation module: for each operator, it allows to annotate or associate a trust degree to each query. This module is used by statistics generator module to calculate the rate of trust and number of trusted or untrusted queries.

$
\scriptsize{\textbf{TrustDegree \ =}}\left\{
\begin{array}{l}
\scriptsize{Boolean \ value: \ 0 \ \Leftrightarrow \ Q \in UTrustQ \ ; 1 \ \Leftrightarrow \ Q \in \ TrustQ} \\
\scriptsize{Categorical\  value } \end{array}
\right.$

\item Loader module: used to store cleansed queries in the specified destination, even database (SQLite in our case) or files. Other destinations will be considered in next versions.

\item Statistics generator module: allows calculating the necessary metrics like Rate of trust and number of Trusted \& Untrusted queries at each step basing on the results of trust annotation module. The resulted metrics’ values for each operator are stored in Yaml files.
\end{enumerate}

\subsection{Data Layer}: 
In this layer, SQL lite Database and file structure are used to store data, either at the end of the pipeline (load operation), or the intermediate resulted curated queries after each operation. Jena TDB\footnote{\url{https://jena.apache.org/documentation/tdb/}}  triple store is used to store LOD ontologies.

The source code is available on: \\\url{https://github.com/dihiaselma/TrustETL} 

The source code is available on: \url{https://github.com/dihiaselma/TrustETL}

\begin{figure}[h]
		\centering
		\includegraphics[scale=0.4]{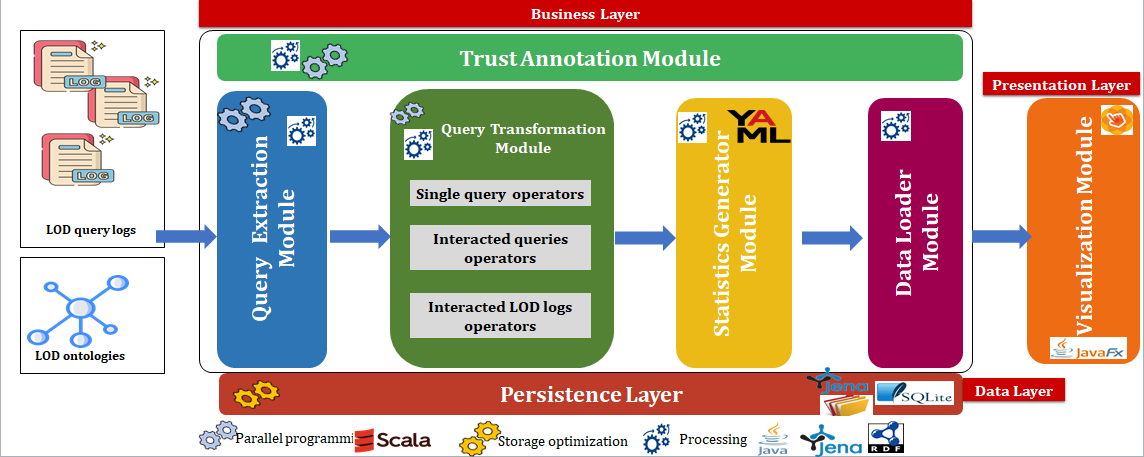}
		\caption{Trust Curator tool Architecture} 
		\label{architecture}
				
\end{figure}

\section{Demonstration Overview}

For the demonstration of Trust curation tool, we simulate the process of curating ScholarlyData logs and DBpedia query logs (we filtered queries related to research topic). These logs are provided by LSQ\footnote{https://aksw.github.io/LSQ/}.
The series of experiments are executed on a machine OS Windows 10x64 with 6 GB RAM and Intel R core TM i7-3632QM, @ 2.20 GHz CPU.  
Scholarly data log contains 5.499.797 raw queries (SPARQL and GET/SET queries) while the DBpedia log contains 3.193.672 SPARQL raw queries, 6.680 queries after research topic filter.

\begin{figure}[h]
		\centering
		\includegraphics[scale=0.7]{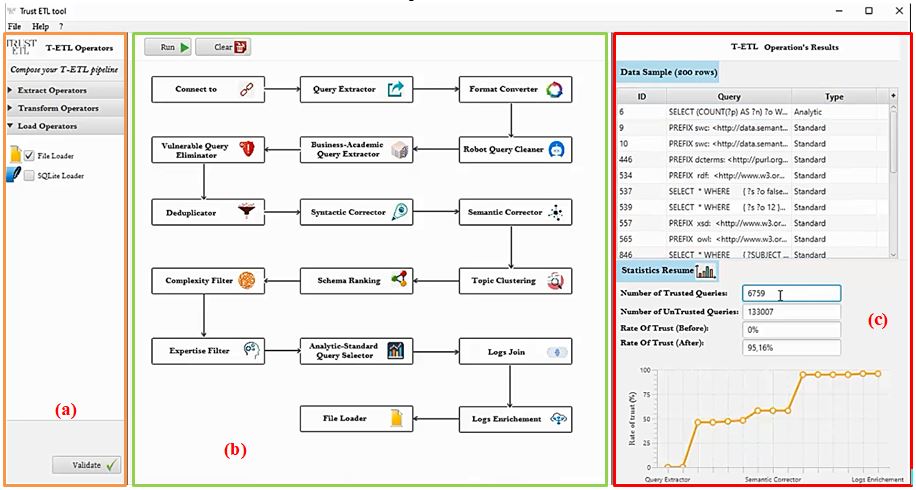}
		\caption{Trust Curator tool GUI} 
		\label{tool}
					
\end{figure}
To curate these logs, we select all necessary operators proposed in zone (a, figure\ref{tool}), the obtained pipeline is shown in zone (b, figure\ref{tool}). Once executed, the resulted metrics and defined statistics are returned in zone (c, figure\ref{tool}).
To get more details, we can click on one operator in zone (b, figure\ref{tool}) and the related statistics are displayed in zone (c, figure\ref{tool}).
The process starts by connecting to scholarly data log file, extracting select \& construct queries and parsing them to UTF8 to obtain 139.932 queries.  Then, series of transformations are applied in a logic order. We start by cleaning robot queries to get 75.467 queries. We can decide to keep just business queries; here we keep all queries. After that, the vulnerable queries are discarded and a deduplication is executed to get 75.100 queries. Semantic \& syntactic errors are widely present, consequently, we proceed to correct them in order to enhance their quality. For topic section, we keep all topics and we proceed to schema ranking in order to clean them form non informative queries, we obtain 6.700 queries. The operation of complexity filter is used to detect shapes and depths of queries helping in next step to select the profile to keep. In our case, we keep all profiles and all types of queries (analytic or standard). At the end, we enrich the scholarly data logs by some queries of DBpedia logs by selection semantically similar queries. The curated trusted queries are loaded into a file. The T-Curator enhances the Rate of trust from 79\,\% to reach 95,16\,\%, and the number of queries is decreased from 139.932 to 6.756 trusted queries.
The video Demo is available on: \url{https://www.youtube.com/watch?v=u25CIUVG0X8}

\section{Conclusion}
In this paper, we presented a new tool Trust Curator that assists Data analyst to curate her LOD logs. The Trust Curator is a tool allowing to combine many proposed trust based operators generating a curation pipeline with some statistics and metrics. This tool permits to enhance the trust of LOD logs since they suffer from many risks linked to their provenance and quality. 

\section{Acknowledgments}
We extend our gratitude to Professor BELLATRECHE Ladjel and Dr. KHOURI Selma for their invaluable guidance, insightful ideas, and contributions.

\bibliographystyle{unsrtnat}
\bibliography{references}

\end{document}